\documentclass[pra,aps,twocolumn]{revtex4}

\usepackage{graphicx,amsmath}
\usepackage{txfonts}
\usepackage{hyperref}

\newcommand{\rp}[1]{(\ref{#1})}

\newcommand{\abs}[1]{\left|{#1}\right|}

\newcommand{\av}[1]{\left\langle #1 \right\rangle}

\newcommand{\al}[1]{^{(#1)}}
\newcommand{\da}{^\dagger}

\newcommand{\pt}[1]{\left( #1 \right)}
\newcommand{\pq}[1]{\left[ #1 \right]}
\newcommand{\pg}[1]{\left\{ #1 \right\}}

\newcommand{\lpg}[1]{\left\{ #1 \right.}

\newcommand{\rpg}[1]{\left. #1 \right\}}
\newcommand{\ee}{{\rm e}}
\newcommand{\ii}{{\rm i}}
\newcommand{\dd}{{\rm d}}

\newcommand{\nn}{{\nonumber}}

\begin{document}

\title{
Suppression of Stokes scattering and improved optomechanical cooling with squeezed light
}

\author{Muhammad~Asjad, Stefano~Zippilli, David~Vitali}
\affiliation{Physics Division, School of Science and Technology, University of Camerino, via Madonna delle Carceri, 9, I-62032 Camerino (MC), Italy, and INFN, Sezione di Perugia, Italy}
\date{\today}

\begin{abstract}
We develop a theory of optomechanical cooling with a squeezed input light field. We show that Stokes heating transitions can be \emph{fully} suppressed when the driving field is squeezed below the vacuum noise level at an appropriately selected squeezing phase and for a finite amount of squeezing.
The quantum backaction limit to laser cooling can be therefore moved down to zero and the resulting final temperature is then solely determined by the ratio between the thermal phonon number and the optomechanical cooperativity parameter, independently of the actual values of the cavity linewidth and mechanical frequency. Therefore driving with a squeezed input field allows to prepare nanomechanical resonators, even with low resonance frequency, in their quantum ground state with a fidelity very close to one.
\end{abstract}

\maketitle

\section{Introduction}

Mechanical resonators hold promise for the development of novel quantum devices that make use of quantum mechanics to achieve enhanced performances for sensing, metrology, storage and transduction of information, and possibly to explore the validity of quantum mechanics at the macroscopic scale~\cite{Genes,Poot,Chen,Aspelmeyer}. In order to operate a mechanical resonator at the quantum level, and overcome the detrimental effects of thermal noise, it has to be cooled to the ground state of motion. Many approaches have been discussed including feedback schemes~\cite{Mancini98,Hamerly12,Wilson15}, cavity-assisted approaches~\cite{Wilson-Rae07,Marquardt07,Genes08}, the coupling with artificial atoms and with spins~\cite{Wilson-Rae04,Rabl09}, the use of electrons in place of photons~\cite{ZippilliPRL09,Sonne10} and the application of coherent control techniques~\cite{Wang11}. Cavity sideband-cooling is one of the most promising approaches, already realized in a number of experiments~\cite{Schliesser08,Schliesser09,Park,Rocheleau,Teufel11,Chan,Karuza,Peterson16,Teufel16}. It consists in the engineering, by means of laser light, of an effective low temperature bath for the mechanical excitations. The effective bath compete with the natural thermal environment to determine the final temperature of the mechanical mode which can be expressed in terms of the steady state number of excitations $N_{st}$ as~\cite{Aspelmeyer,Wilson-Rae07,Marquardt07,Genes08}
\begin{eqnarray}\label{Nst0}
N_{st}=\frac{\gamma\,N_{th}+\Gamma\,N_a}{\gamma+\Gamma}\ ,
\end{eqnarray}
where $N_{th}$ is the number of excitations corresponding to the natural thermal reservoir, $\gamma$ is the dissipation rate into the thermal bath, $N_a$ is the quantum back-action limit, i.e., the effective number of excitations of the thermal bath realized by the light, and $\Gamma$ is the cooling rate, i.e., the corresponding light-induced dissipation rate.
Cooling is achieved by engineering fast dissipation ($\Gamma \gg \gamma $) into an effective low temperature bath ($N_a\ll N_{th}$).
The values of $\Gamma$ and $N_a$ are determined by the response of the mechanical resonator to the incident light, and more specifically by the scattering rate of light at the Stokes ($A_+$) and anti-Stokes ($A_-$) sidebands, corresponding to the increase and decrease of one mechanical energy quantum, respectively. Hence, cooling is achieved when the laser cooling rate $\Gamma=A_--A_+$ is positive, while the back-action limit is $N_a=A_+/\Gamma$.
In cavity-optomechanics the unbalance between Stokes and anti-Stokes processes is obtained by means of an optical cavity. The resonator interacts with the cavity photons which are pumped by a laser drive, red-detuned with respect to the relevant cavity resonance; the anti-Stokes scattered photons are hence made resonant,  and cooling takes place. However, the residual non-resonant Stokes heating processes sets a fundamental limit to the achievable occupancy: in fact, in the optimal case $\Gamma \gg \gamma $ and not too large temperature $\gamma N_{th}/\Gamma < N_a$, one achieves the quantum back-action limit $N_{st}\sim N_a$, which is determined by the effective temperature of the light-induced thermal bath. In standard laser cooling the non-resonant Stokes scattering is minimized using a narrow cavity linewidth $\kappa_a$, much smaller than the mechanical frequency $\omega_m$, thus entering the so called resolved sideband regime, where
$N_a\geq\kappa_a^2/4\omega_m^2$. This limit is observed also in atomic laser cooling under specific conditions, and it has been suggested to overcome it by engineering quantum destructive interference processes with multilevel-systems~\cite{Morigi00,Evers04,ZippilliPRL05,ZippilliPRA05}. In optomechanics it has been proposed to realize a similar destructive interference by exploiting the effect of optomechanical induced transparency~\cite{Ojanen} and by engineering cavity dissipation modulated by the resonator position~\cite{Elste09,Sawadsky15}.

Here we demonstrate that the \emph{full} suppression of Stokes scattering (corresponding to $A_+=0$, and hence to an effective zero-temperature light-induced thermal reservoir with $N_a=0$) can be achieved when an optomechanical system is driven by a light field squeezed below the vacuum noise level by a finite amount and at well defined squeezing phase. One has to consider an experimental setup very close to that realized in the experiments reported in Refs.~\cite{Clark,Schafermeier,Clark2016}, where the squeezed light is generated by parametric amplification. Correspondingly it can be described in the framework of cascaded open quantum systems~\cite{cascade,Asjad16}. Squeezed light is the fundamental resource in quantum information with continuous variables and in many application of quantum metrology~\cite{Weedbrook,ZippilliNJP15}. A number of recent experiments have started to explore the potentiality of squeezed light for the manipulation of quantum systems~\cite{Clark,Schafermeier,Clark2016,Murch,LIGO13,Steinlechner,Baune,Lucivero,Taylor}, and various proposals suggested to generate nonclassical states of mechanical systems through the injection of squeezed light~\cite{Zhang03,Jahne,Asjad16,ZippilliPRA15}.

The use of squeezed light driving for improving cooling has been proposed in Ref.~\cite{Cirac93} in the case of the motion of a trapped ion. However, in Ref.~\cite{Cirac93}, cooling to the ground state is achieved only for the unrealistic limit of an infinitely squeezed field, and in this limit the cooling rate tends to zero, i.e., cooling becomes extremely slow. On the contrary, in the optomechanical case studied here optimal cooling is achieved with \emph{finite} squeezing, and the cooling rate remains unchanged, as a result of destructive quantum interference due to the scattering of the correlated photons of the squeezed field.

\section{The System}

The system can be described in terms of linearized quantum Langevin equations for the annihilation and creation operators of cavity photons, $a$ and $a\da$, and of mechanical excitations, $b$ and $b\da$, given by~\cite{Genes} $\dot{a}= -\pt{\kappa_{a}+ \ii\Delta_{a}}\, a+\ii\,G\,\pt{b+b\da},+\sqrt{2\kappa_a}\, a_{in}\,$ and $\dot{b}=-\pt{\ii\omega_m+\frac{\gamma}{2}}b+\ii\,G \pt{a^\dagger+ a}+\sqrt{\gamma_j}\,b_{in}, \label{linQLE}\,$ where $\Delta_a$ is the detuning between the relevant cavity mode and the driving laser field, $G$ is the linearized optomechanical coupling strength, and $b_{in}(t)$ is the delta correlated mechanical noise operators which accounts for the mechanical effects of the thermal environment at temperature $T$, such that $\pq{b_{in}(t),b\da_{in}(t')}=\delta(t-t')$ and $\langle b_{in}(t) b_{in}\da(t')\rangle =(N_{th}+1)\delta(t-t')$, with  $N_{th}=(\ee^{\hbar \omega_{m}/K_B T}-1)^{-1}$.  Finally the noise operator $a_{in}$ accounts for the effect of the external electromagnetic environment. It describes a squeezed reservoir whose properties are determined by the output light of the parametric oscillator (with annihilation operator $c_{out}\al{s}$) which drives the system in a cascade configuration. The corresponding description is based on the theory of open quantum cascade systems developed in Refs.~\cite{cascade} (a similar model have been discussed in detail in Ref.~\cite{Asjad16}).
Specifically $a_{in}$ can be written as
$a_{in}=\frac{1}{\sqrt{\kappa_a}}\pt{\sqrt{\kappa_a\al{s}}\,c_{out}\al{s}+\sqrt{\kappa_a'}\,a_{in}'}$ ,
where we have decomposed it as the sum of two uncorrelated bosonic operators: $c_{out}\al{s}$ for the squeezed reservoir, which exchanges photons with the cavity at rate $\kappa_a\al{s}$, and $a_{in}'$ for residual vacuum modes of the electromagnetic environment into which the cavity can decay at rate $\kappa_a'$ (optical losses), with $\kappa_a=\kappa_a\al{s}+\kappa_a'$. 
The residual vacuum modes are characterized by the correlation function $\av{a'_{in}(t)\ {a'_{in}}\da(t')}=\delta(t-t')$, while the squeezed reservoir by 
$\av{c_{out}\al{s}(t)\,{c_{out}\al{s}}\da(t')}=\delta(t-t')+n(t-t')$ 
and
$\av{c_{out}\al{s}(t)\,c_{out}\al{s}(t')}=m(t-t')$, 
where we have introduced the functions $n(\tau)$ and $m(\tau)$, whose specific form is given below. 
They determine, respectively the number of excitations and the strength of the field self-correlations, and can be expressed in terms of the parameters of the parametric oscillator, namely the non-linear self-interaction strength $\chi$, and the linewidth of the optical resonator $\kappa_c$, such that the variables of the parametric oscillator (annihilation and creation operators $c$ and $c\da$) fulfill the equation 
$\dot c=-\kappa_c\,c+\chi\,c\da+\sqrt{2\kappa_c}c_{in}\ .$
Here the optical mode of the parametric oscillator is resonant with the laser field which drives the optomechanical system.
The input noise operator fulfills the relation $\av{c_{in}(t)\,c_{in}\da(t')}=\delta(t-t')$, and also in this case can be decomposed as 
$c_{in}=\frac{1}{\sqrt{\kappa_c}}\pt{\sqrt{\kappa_c\al{s}}\,c_{in}\al{s}
+\sqrt{\kappa_c'}\,c_{in}'}\ ,$
with $\kappa_c=\kappa_c\al{s}+\kappa_c'$, where $c_{in}\al{s}$ corresponds to the external modes of the electromagnetic field which are controlled and used to drive the optomechanical system, while $c_{in}'$ accounts for residual uncontrolled modes;
furthermore the output field  fulfill the standard relation
 $c_{out}\al{s}=\sqrt{\kappa_c\al{s}}\,c-c_{in}\al{s}$. In detail we find
\begin{eqnarray}\label{nm}
n(\tau)&=&\frac{\chi\,\kappa_c\al{s}}{2}
\pq{\frac{ \ee^{-r_-\abs{\tau}}}{r_-}-\frac{ \ee^{-r_+\abs{\tau}}}{r_+}
}
\nn\\
m(\tau)&=&\frac{\chi\,\kappa_c\al{s}}{2}
\pq{\frac{ \ee^{-r_-\abs{\tau}}}{r_-}+\frac{ \ee^{-r_+\abs{\tau}}}{r_+}
}\,\ee^{-2\ii\,\phi} 
\ ,
\end{eqnarray}
where $r_\pm=\kappa_c\pm\chi$ and $\phi$ is the phase of squeezing.
In particular $r_+$ is the decay rate of the fluctuations of the maximum squeezed quadrature of the output field, that in this case is $Y_{out}\al{s}=c_{out}\,\ee^{\ii(\pi/2+\phi)}+{c_{out}\al{s}}\da\,\ee^{-\ii(\pi/2+\phi)}$ (namely $r_+$ is the squeezing bandwidth). Instead $r_-$ is the decay rate of the fluctuations of the anti-squeezed quadrature.

It is convenient to consider the correlation functions for the total noise operator for the optical cavity $a_{in}$. They are given by $\av{a_{in}(t)\,{a_{in}}\da(t')}=\delta(t-t')+n_\xi(t-t')$ and 
$\av{a_{in}(t)\,a_{in}(t')}=m_\xi(t-t')\,\ee^{-2\ii\,\phi}$ with
\begin{eqnarray}\label{nm}
n_\xi(\tau)=\xi\,n(\tau) \hspace{1cm} {\rm and} \hspace{1cm}  m_\xi(\tau)=\xi\,m(\tau)
\end{eqnarray}
where we have introduced the scaling factor $\xi=\kappa_a\al{s}\kappa_c\al{s}/\kappa_a\kappa_c$ which accounts for possible uncontrolled dissipation channels whose effect is that of reducing the purity of the driving squeezed field. In particular this model describes a thermal squeezed bath, with pure squeezing obtained for $\xi=1$ when no uncontrolled optical losses are taken into account. In the opposite limit of maximum loss, $\xi=0$, only standard optical vacuum noise enters the cavity. 
Correspondingly the spectra of correlations of the input field operators [i.e. the Fourier transform of $n_\xi(t)$ and $m_\xi(t)$ which are used below in the expressions for the Stokes and anti-Stokes scattering rates] are given by 
\begin{eqnarray}
\tilde n_\xi(\omega)&=&\xi\,\chi\, \kappa_c\pt{\frac{1}{r_-^2+\omega^2}-\frac{1}{r_+^2+\omega^2}}
\nn\\
\tilde m_\xi(\omega)&=&\xi\,\chi\,\kappa_c\pt{\frac{1}{r_-^2+\omega^2}+\frac{1}{r_+^2+\omega^2}}\ .
\nn
\end{eqnarray}

\section{Cooling}

\begin{figure}[t!]
\includegraphics[width=4.2cm]{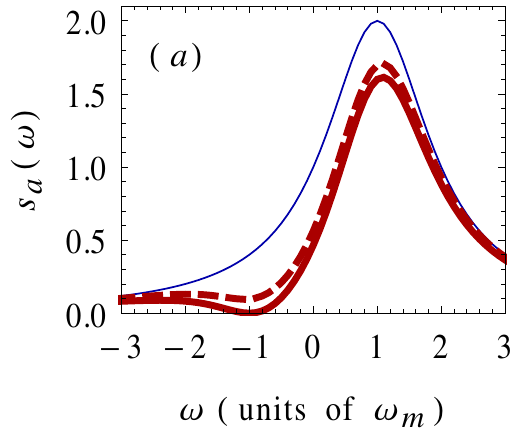}
\includegraphics[width=4.2cm]{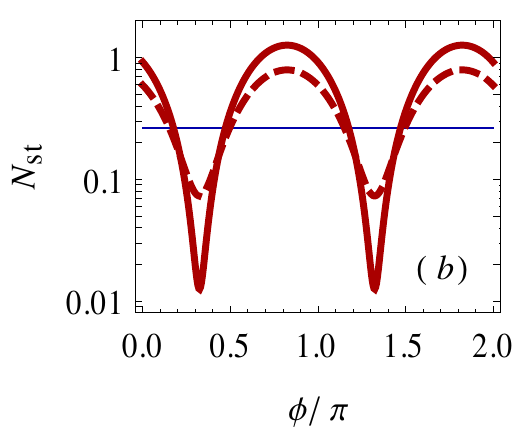}
\caption{
(a) Power spectrum of the radiation pressure force $s_a(\omega)$ for parameters which fulfill Eqs.~\rp{phi} and \rp{cond1}. The values of these curves at $\omega=\pm\omega_m$ determine the rates for Stokes and anti-Stokes scattering.
(b) Steady state excitation number $N_{st}$ as a function of the squeezing phase, for parameters that fulfill Eqs.~\rp{cond1}. The lines in both plots are evaluated for a driving field squeezed by 5 dB below the vacuum noise level at the central frequency, i.e. $S(0)=0.3$, and for (thick solid red line) $\xi=1$, (thick dashed red line) $\xi=0.8$ and (thin blue line) $\xi=0$ (no squeezing). In (a) $\phi=0.3\pi$. The other parameters are $\Delta_a=\omega_m$, $\kappa_a=\omega_m$, $G=0.1\omega_m$, $\gamma=0.2\times10^{-6}\omega_m$, $N_{th}=1000$.
}\label{fig1}
\end{figure}

When $\gamma\ll G\ll r_\pm,\kappa,\omega_m$ the dynamics of the mechanical resonator can be approximated by eliminating the cavity degrees of freedoms at the lowest relevant order in the coupling parameter $G$. The resulting equation for the average number of mechanical excitations $N(t)=\av{b\da(t)\,b(t)}$ is given by 
\begin{eqnarray}
\dot N(t)=-\pt{\gamma+\Gamma}\,N(t)+\gamma\,N_{th}+A_+\ ,
\end{eqnarray}
so that the steady state is given by Eq.~\rp{Nst0}.
The lowest order expressions for the Stokes and anti-Stokes scattering rates are given by $A_\pm=G^2\,s_a(\mp\omega_m)$, where $s_a(\omega)=\int_{-\infty}^\infty\dd t\,\ee^{\ii\omega t}\av{Y(t)\,Y(0)}_{st}$ is the spectrum of fluctuations of the cavity field quadrature which couples to the mechanical resonator $Y=a+a\da$ (the spectrum of fluctuations of the force operator~\cite{Aspelmeyer}), and where the correlation function is evaluated in the steady state and at zeroth order in the optomechanical coupling $G$ (namely it is evaluated in the steady state of an empty cavity driven by a squeezed field).
It is explicitly given by
\begin{eqnarray}\label{S}
s_a(\omega)&=&\frac{2\kappa_a}{\kappa_a^2+\pt{\Delta_a-\omega}^2}
\lpg{1+
\tilde n_\xi(\omega)\pq{1+\frac{\kappa_a^2+\pt{\Delta_a-\omega}^2}{\kappa_a^2+\pt{\Delta_a+\omega}^2}}
}\nn\\&&\hspace{-0.9cm} \rpg{
-2\,\tilde m_\xi(\omega)\,\frac{
\pt{\Delta_a^2-\kappa_a^2-\omega^2}\cos\pt{2\phi}+
2\kappa_a\,\Delta_a\,\sin\pt{2\phi}
}{\kappa_a^2+\pt{\Delta_a+\omega}^2}
}\ .
\end{eqnarray}
We first note that the value of $\Gamma$ does not depend on the squeezed light, and it is actually equal to the standard optomechanical cooling rate
$\Gamma=A_--A_+=2\kappa_a\,G^2\pq{\frac{1}{\kappa_a^2+\pt{\Delta_a-\omega_m}^2}-\frac{1}{\kappa_a^2+\pt{\Delta_a+\omega_m}^2}}$~\cite{Wilson-Rae07,Marquardt07,Genes08}. The largest cooling rate is therefore achieved in the resolved sideband regime, $\kappa\ll\omega_m$, and when the laser is set at the red mechanical sideband frequency $\Delta_a=\omega_m$~\cite{Wilson-Rae07,Marquardt07,Genes08}. However, more importantly, the back-action limit $N_a$ can be strongly affected by the squeezed light. In particular the term proportional to $\tilde m_\xi(\omega)$ in Eq.~\rp{S} can be made negative and can contribute to the suppression of the light-induced heating processes. The power spectrum $s_a(\omega)$ may become like the one in Fig.~\ref{fig1}(a) when particular conditions on the phase and on the amount of squeezing are satisfied as specified below. It displays the characteristic Fano-like profile typical of interference phenomena. In this case it is due to the interference between processes involving the exchange of photons between the squeezed reservoir and the squeezed cavity field. In detail, processes in which the increase of a mechanical excitation is accompanied by the annihilation or creation of a cavity photon and subsequent emission or absorption of a photon to or from the reservoir, can interfere destructively depending on the squeezing phase. As a result, the value of $s_a(\omega)$ at $-\omega_m$, which determines the strength of the Stokes heating transitions $A_+$, can be completely suppressed. At the same time also the peak value at $\omega_m$ for the anti-Stokes cooling transitions is reduced as compared to the standard result (thin blue line), such that their difference $s_a(\omega_m)-s_a(-\omega_m)$ (proportional to the cooling rate $\Gamma$) remains unaffected by the squeezed light. We note that perfect suppression of Stokes scattering can be observed only when $\xi=1$, that is, when no uncontrolled optical losses are present; however, strong reduction is observed also for reasonable values $\xi<1$, as described by the dashed line in Fig.~\ref{fig1}(a) which includes $20\%$ of uncontrolled optical losses.

\section{Results}

We now optimize the cooling by minimizing $s_a(-\omega_m)$, and hence the back-action limit $N_a$. As a function of the squeezing phase relative to the phase of the pump field, such a minimum is obtained when (for $\Delta_a>0$)
\begin{eqnarray}\label{phi}
\phi=\frac{1}{2}\arctan\pt{\frac{2\Delta_a\kappa_a}{\Delta_a^2-\omega_m^2-\kappa_a^2}} +k\pi
\end{eqnarray}
and $2k\pi<2\phi<3k\pi$ with $k\in \mathbb{Z}$.
Under this condition we find
\begin{eqnarray}\label{Na1}
N_a=N_0\pq{1+\tilde n_\xi(\omega_m)\pt{1+\frac{1}{\zeta^2}}-2\frac{\tilde m_\xi(\omega_m)}{\zeta}},
\end{eqnarray}
where we have introduced $N_{0}=\pq{\kappa_a^2+\pt{\Delta_a-\omega_m}^2}/\pt{4\,\Delta_a\,\omega_m}$ which is the steady state back-action limit without the squeezed light~\cite{Wilson-Rae07,Marquardt07},
and the parameter $\zeta={\pq{\kappa_a^2+\pt{\Delta_a-\omega_m}^2}^{1/2}\pq{\kappa_a^2+\pt{\Delta_a+\omega_m}^2}}^{-1/2},$ which is smaller than one for $\Delta_a>0$. Then one has to minimize Eq.~(\ref{Na1}) over the properties of the squeezed input driving, i.e., over $\tilde n_\xi(\omega_m)$ and $\tilde m_\xi(\omega_m)$, yielding
\begin{equation}\label{cond1}
N_a^{\rm opt} =N_0(1-\xi),\;\; {\rm for} \;\;\zeta=\frac{\tilde n_\xi(\omega_m)}{\tilde m_\xi(\omega_m)}.
\end{equation}
In the absence of squeezing, $\xi=0$, we recover the standard result of laser sideband cooling~\cite{Wilson-Rae07,Marquardt07}, while, as anticipated, the quantum back-action limit is \emph{fully} suppressed when $\xi=1$, i.e., when only pure squeezed light and no vacuum noise enters the cavity.

Under these optimized conditions for the squeezed input, one has a significant suppression of the phonon occupancy $N_{st}$ of Eq.~(\ref{Nst0}), as shown in Fig.~\ref{fig1}(b) where $N_{st}$ is plotted versus the squeezing phase. It displays minima with periodicity $\pi$ [see Eq.~\rp{phi}] and $N_{st}$ can drop well below the result achieved with standard sideband cooling (thin blue line). The importance of the full suppression of the back-action limit achievable with the squeezed driving can be seen also from Eq.~(\ref{Nst0}), which in this optimal case, and in the usual conditions $\gamma\ll G\ll r_\pm,\kappa,\omega_m$, and red sideband driving $\Delta = \omega_m$, can be rewritten as
\begin{equation}\label{Nstapp}
   N_{st}\simeq \frac{N_{th}}{C} +N_0(1-\xi),
\end{equation}
where $C=2G^2/\gamma\kappa_a$ is the cooperativity; taking $\xi=1$ and very large $C/N_{th}$, $N_{st}$ can become arbitrarily small and the mechanical resonator can be prepared in its quantum ground state with very high fidelity.

\begin{figure}[t!]
\includegraphics[width=4.4cm]{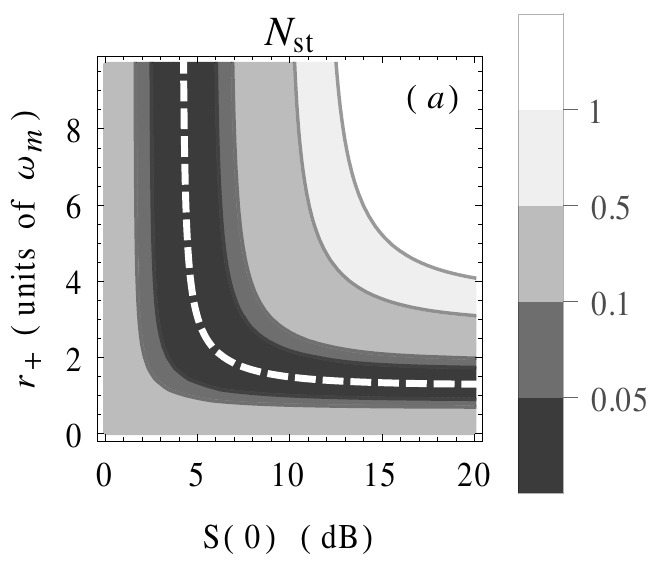}
\hspace{-0.3cm}
\includegraphics[width=4.2cm]{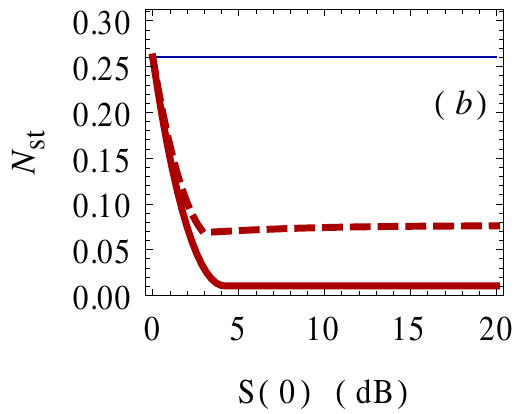}
\caption{Steady state mechanical excitation number $N_{st}$ as a function of the value of squeezing at the central frequency $S(0)$ and of the squeezing bandwidth  $r_+$, for a value of the squeezing phase that fulfill Eq.~\rp{phi}, $\phi=0.3\,\pi$. In (a) $\xi=1$ and the dashed curve corresponds to the parameters that fulfill Eq.~\rp{cond2}. The curves in (b) are evaluated for the values of $r_+$ that fulfill Eq.~\rp{cond2}, i.e. it is evaluated along the dashed curve in (a).
The other parameters and the line styles are as in Fig.~\ref{fig1}.
}\label{fig2}
\end{figure}

It is convenient to re-express the optimal condition for the squeezed input field of Eq. (\ref{cond1}) in terms of easily measurable quantities. We introduce  the input squeezing spectrum $S(\omega)=\int_{-\infty}^\infty\dd t\,\ee^{\ii\omega t}\av{Y_{in}(t)\,Y_{in}(0)}$, namely the power spectrum of the maximally squeezed quadrature of the output of the parametric amplifier (and cavity input as well), that in this case is $Y_{in}=a_{in}\,\ee^{\ii(\pi/2+\phi)}+{a_{in}}\da\,\ee^{-\ii(\pi/2+\phi)}$. The field is squeezed below the vacuum noise level when the condition $S(\omega)<1$ is satisfied. In the present case it is $S(\omega)=1+2\tilde n_\xi(\omega)-2\tilde m_\xi(\omega)$ and maximum squeezing [i.e., the minimum of $S(\omega)$] is observed at the central frequency $\omega = 0$ and it extends over a bandwidth $r_+$. It is possible to rewrite the parameters $\tilde n_\xi(\omega_m)$ and $\tilde m_\xi(\omega_m)$ in terms of the value of the corresponding squeezing spectrum at the central frequency, $S(0)$, and of the squeezing bandwidth $r_+$, which can be both easily measured. In particular we find that the optimal cooling condition in Eq.~\rp{cond1} can be rewritten as
\begin{eqnarray}\label{cond2}
\frac{\omega_m^2}{r_+^2}=\pq{1-S(0)}\frac{1+\zeta}{2\xi\,\zeta}-1,
\end{eqnarray}
which relates the amount of squeezing to the bandwidth, and can be exactly fulfilled only if $S(0)\leq 1-2\xi\,\frac{\zeta}{1+\zeta}$. In particular the smaller the amount of squeezing the larger the bandwidth needed to obtain perfect Stokes scattering suppression. Instead, in the case in which the field is not sufficiently squeezed ($S(0)> 1-2\xi\,\frac{\zeta}{1+\zeta}$), the best condition is obtained only in the limit of an infinite bandwidth $r_+\to \infty$, where however the Stokes heating transitions are not completely suppressed, and $N_{a}=N_0\pg{1+\frac{1-S(0)}{S(0)-1+\xi}\pq{\frac{1-S(0)}{4}\pt{1+\frac{1}{\zeta}}^2-\frac{\xi}{\zeta}}}$. 

The behavior of the steady state number of excitations $N_{st}$ as a function of the squeezing parameters $S(0)$ and $r_+$, and for a squeezing phase fixed at the optimal value of Eq.~\rp{phi}, is reported in Fig.~\ref{fig2}. Plot (a) clearly shows a strong reduction of $N_{st}$ in the parameter region around the dashed line corresponding to the optimal condition of Eq.~\rp{cond2}. 
The minimum value of $N_{st}$ as a function of $S(0)$, when $r_+$ is varied in order to fulfill Eq.~\rp{cond2} [namely the value of $N_{st}$ along the dashed curve in Fig.~\ref{fig2} (a)], is instead reported in plot (b) where the solid and dashed thick lines correspond to $\xi=1$ and $\xi<1$ respectively. We note that the fast increase of $N_{st}$ for small squeezing corresponds to values for which Eq.~\rp{cond2} cannot be exactly satisfied. In any case, one has an improvement with respect to the corresponding value of $N_{st}$ achieved without the squeezed field (thin blue line); moreover, when the condition of Eq.~\rp{cond2} is satisfied (at larger squeezing), the reduction of the steady state mechanical population is independent of the actual value of the squeezing as a result of the suppression of Stokes scattering. What is relevant is that perfect Stokes scattering suppression is always achieved at \emph{finite} squeezing; for smaller squeezing, full suppression requires only a larger bandwidth. This is not possible in the trapped ion case studied in Ref. \cite{Cirac93}, where this is achievable only in the limit of infinite squeezing.

\begin{figure}[t!]
\includegraphics[width=4.4cm]{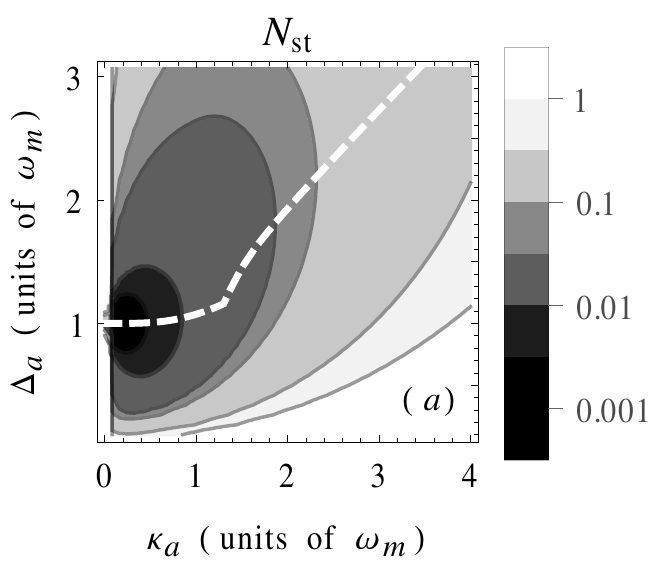}
\hspace{-0.3cm}
\includegraphics[width=4.4cm]{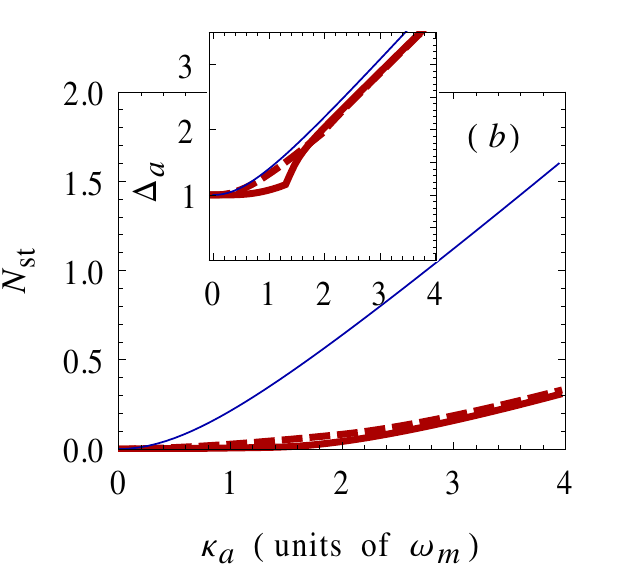}
\caption{
Steady state mechanical excitation number $N_{st}$ as a function of the cavity decay rate $\kappa_a$ and detuning $\Delta_a$,
for values of $\phi$ and $r_+$ that fulfill Eqs.~\rp{phi} and \rp{cond2}. In (a) $\xi=1$ and the dashed curve corresponds to the values of $\Delta_a$ which minimize $N_{st}$ at each value of $\kappa_a$. The lines (b) are evaluated for the values of $\Delta_a$ which minimize $N_{st}$ at each value of $\kappa_a$,  The specific values of $\Delta_a$ for each line are reported in the inset [the thick solid red line in the inset is equal to the dashed line in (a)]. The other parameters and the line styles are as in Fig.~\ref{fig1}.
}\label{fig3}
\end{figure}

In Fig.~\ref{fig3} we study the performance of this cooling protocol against the optomechanical parameters, the cavity decay rate $\kappa_a$ and the detuning $\Delta_a$, with the squeezed light chosen in order to minimize the Stokes scattering. Specifically these results are evaluated at fixed $S(0)$ and for values of $r_+$ which fulfill Eq.~\rp{cond2} when $\kappa_a$ is sufficiently small, and for $r_+\to\infty$ when $\kappa_a$ is so large that Eq.~\rp{cond2} cannot be fulfilled.
The dashed line in Fig.~\ref{fig3} (a) indicates the values of $\Delta_a$ for which $N_{st}$ is minimum for each value of $\kappa_a$. The kink visible in this line divides the range of parameters for which the condition in Eq.~\rp{cond2} can be fulfilled exactly (small $\kappa_a$) from that at which it can not. The specific minimum values of $N_{st}$ along this line are instead reported in Fig.~\ref{fig3} (b) (thick solid red line) where it is compared with the results evaluated at finite uncontrolled optical losses (thick dashed red line) and with the standard laser cooling without squeezing (thin solid  blue line). Although the minimum is achieved as expected at $\Delta_a=\omega_m$ and small $\kappa_a$, a wide range of very low temperature is observed for parameters which extend far beyond the regime of resolved sideband cooling $\kappa_a\ll\omega_m$. 
We find for example that a realistic mechanical resonator with quality factor $Q=5\times 10^{6}$ , resonant frequency $\omega_m=2$MHz and interacting with the field of an optical cavity with linewidth $\kappa_a=\omega_m$, can be cooled from a temperature of $T=0.1$K ($N_{th}\sim1000$) to the final temperature of $T_{final}=0.02$mK corresponding to $N_{st}=0.01$ mechanical excitations, when it is driven by a field squeezed by 5dB below the vacuum noise level over a bandwidth of $r_+=3\,\omega_m$. Under the same condition, standard sideband cooling would give $N_{st}=0.26$. We note that, in this specific example corresponding to already demonstrated technologies, the $99\%$ ground state fidelity, achievable with the squeezed field, could be of fundamental importance for the implementations of quantum information protocols or for fundamental tests of quantum theories, which requires ground state cooling and which could not tolerate the $26\%$ of error in the steady state preparation corresponding to standard laser cooling.

\section{Conclusions}

In conclusion we have studied the cooling dynamics of a mechanical resonator coupled by radiation pressure to a resonant mode of an optical cavity driven by a squeezed field in a configuration similar to that investigated in recent experiments~\cite{Clark,Schafermeier,Clark2016}.
The interplay between squeezed field and mechanical vibrations can lead to the complete suppression of Stokes scattering
by quantum destructive interference,
when the squeezing phase and the amount of squeezing at the mechanical sideband frequency are appropriately selected. 

Injecting squeezed vacuum light in the cavity allows to beat, by using only a \emph{finite} amount of squeezing, the typical constraints of optomechanical ground state cooling. In particular the requirement of being deep in the resolved sideband regime is no more necessary, and the quantum backaction limit, that defines the ultimate efficiency for standard sideband cooling, can be brought to zero, in the case of pure squeezed driving. This makes the presented protocol very relevant and attractive for approaching the, yet elusive, ground state cooling of low frequency resonators which is necessary for the investigation of macroscopic quantum phenomena.

We note that similar results and the experimental observation of the enhanced optomecahnical cooling, as predicted here, have been recently reported in Ref.~\cite{Clark2016}.

\section*{Acknowledgments}
D.V. thanks John Teufel for an inspiring discussion that stimulated our interest towards this problem. This work is supported by the European Commission through the Marie Curie ITN cQOM and FET-Open Project iQUOEMS.

\end{document}